\documentclass[11pt]{article}
\usepackage{amsmath}
\usepackage{amsfonts}
\usepackage{amssymb}
\usepackage{graphicx}

\begin{document}
\begin{center}
\LARGE {\bf Brane Cosmology With  Generalized Chaplygin Gas in The Bulk }
\end{center}
\begin{center}
{\bf  Kh. Saaidi}\footnote{ksaaidi@uok.ac.ir},\\
{\bf  A. H. Mohammadi}\footnote{Abolhassan.Mohammadi@uok.ac.ir}\\

{\it Department of Physics, Faculty of Science, University of
Kurdistan,  Sanandaj, Iran}

\end{center}
 \vskip 1cm
\begin{center}
{\bf{Abstract}}
\end{center}
We find exact solution of the Einstein equations in the context
of the brane world scenario. We have supposed a { generalized
chaplygin gas} equation of state for bulk. This study display a
constant energy density and pressure for bulk in late time. It is
shown that our assumptions impose a specific equation of state on
brane. {  In this work, we have obtained a decelerate universe in
early time and late time.} In the end, it is shown that under some
assumption we have equation of state of cosmological constant for
bulk.
 \\


\newpage
\section{Introduction}
There is no experimental evidence that support our universe might posses
 any spacetime dimensions beyond the four dimensions, it is a rather remarkable fact
  that we actually know of no compelling reason as to the number of spacetime
  dimension should be four. Consequently, it is legitimated to consider the possible
  existence of additional dimensions beyond four. In fact attempts at a higher-dimensional
   unification were initiated a long time ago. Investigation on the possibility of
   extra dimensions began with Kaluza-Klein type theory (see e.g. {\cite{1}}).
The Randall-Sundrum model {\cite{2}, \cite{3}} suggests the
universe is identified with a four dimensional hypersurface (a
three-brane) in a five dimensional bulk with negative
cosmological constant (AdS space). In their first model, our
universe is assumed to be negative tension brane; however,
Shiromizu et al. {\cite{4}} have shown that, in this model, the
gravitation is repulsive. In their second model they put a single
positive
 tension and suggest the extra dimension need not be compact.
 It has been shown that {\cite{5}}, generically, the equations governing the
 cosmological evolution of the brane will be different from standard cosmology. The
 important difference is related to the quadratical dependence of new
 Friedmann equation to the brane energy density.  For law energy densities on the brane, and a pure AdS,
 we can recover the standard Freidmann-Robertson-Walker
 cosmology (for review, see {\cite{6}, \cite{7}, \cite{8}}).
 The brane evolution can be discussed either in two system
 of coordinates. The first coordinate system,
 namely a GN coordinate, is very useful for a "brane-based"
 point of view {\cite{5}, \cite{9}}. the second coordinate system is known
 as the five dimensional Schwarzschild-Anti de Sitter metric,
 which relies on a "bulk-based" point of view {\cite{10}, \cite{11}}.\\
  Although in string theory
 the bulk field should correspond in some way to the string theory fields,
 from a phenomenological view point, matter in the bulk can be anything {\cite{12}}.
 In this work, we have tried to solve Einstein equation in
 the brane world scenario for a perfect fluid, that is called
  { generalized chaplygin gas}{\cite{13}}. In the standard cosmology, { generalized chaplygin gas}
  model describe a transition from a universe with dust-like matter
   to an accelerated expanding stage. So, in this paper,
   we have investigated the evolution of a homogeneous isotropic
   brane world that embedded in the bulk filled with { generalized chaplygin gas}.
\par The plan of this paper is the following. Section 2 is
preliminary, so we present some basic equations that are useful
for our investigation. In section 3, we impose a { generalized
chaplygin gas} model on bulk, and we have obtained infinite bulk
energy density in early time an a constant value for bulk energy
density in late time. Also, the assumption imposed a specific
equation of state on brane. Our investigation showed that we have
a decelerate universe in early time and deceleration universe in
late time. In section 4, it is assumed that the energy density of
bulk depends only on time, and the coefficient of fifth
coordinate in metric, be independence on time. It is shown that,
under this assumption, the cosmological constant equation of
state is only equation of state which can describe bulk.

\section{Preliminary}
In this section, we introduce some basic equation for our work.
Suppose that five-dimensional spacetime metric is characterized as:
\begin{equation}
ds^2=-n^2(t,y)dt^2+a^2(t,y)\delta_{ij}dx^idx^j+b^2(t,y)dy^2,
\end{equation}
where $\delta_{ij}$ is a maximally symmetric 3-dimensional metric
($k=-1,0,1$ will {\bf parameterize} the spatial curvature), and $y$ is
the fifth coordinate. The 5-D Einstein equations are as usual
form:
\begin{equation}
G_{\alpha\beta}=R_{\alpha\beta}-\frac{1}{2}R_{\alpha\beta}={\kappa}^2T_{\alpha\beta},
\end{equation}
(note that $\alpha,\beta=0,1,2,3,5$), where $R_{\alpha\beta}$ is
5-D Ricci tensor and R is scalar curvature and $\kappa$ is
related to five-dimensional Newton's constant, $G_{(5)}$.
$T_{\alpha\beta}$ is the total energy-momentum tensor. We
distinguish it into two kind of source: \\
 i)the energy-momentum of bulk:
\begin{equation}
T^{\alpha}_{\phantom{\alpha} \beta}|_{B}=diag\left(
-\rho_{B},P_{B},P_{B},P_{B},P_{5} \right),
\end{equation}
where the energy density $\rho_B$ and pressure $P_B$ and $P_5$ are
independent of fifth coordinate.\\
ii)the energy-momentum of brane. We suppose only homogeneous and
isotropic geometry inside the brane,
\begin{equation}
T^{\alpha}_{\phantom{\alpha}
\beta}|_{br}=\frac{\delta(y)}{b}diag\left(
-\rho_{b},p_{b},p_{b},p_{b},0 \right),
\end{equation}
where energy density $\rho_b$ and pressure $p_b$ are independent
of the position inside the brane. In here, we clearly see that,
by this assumption, $T_{05}=0$; it means that there is no flow of
energy along the fifth dimension. Using Eq.(1) and(2), we get\\
\begin{eqnarray}
G_{00}&=&3 \Bigg\{ \frac{\dot{a}}{a} \left(
\frac{\dot{a}}{a}+\frac{\dot{b}}{b} \right) - \frac{n^2}{b^2}
\left( \frac{a''}{a} + \frac{a'}{a} \left( \frac{a'}{a} - \frac{b'}{b} \right) \right) + k \frac{n^2}{a^2} \Bigg\}, \\
G_{ij}&=&\frac{a^2}{b^2}\gamma_{ij}\Bigg\{ \frac{a'}{a} \left(
\frac{a'}{a}+2\frac{n'}{n} \right) - \frac{b'}{b}
 \left( \frac{n'}{n}+2\frac{a'}{a} \right) +2\frac{a''}{a}+\frac{n''}{n} \Bigg\} {} \nonumber \\
 & & + \frac{a^2}{n^2}\gamma_{ij} \Bigg\{ \frac{\dot{a}}{a} \left( -\frac{\dot{a}}{a}+2\frac{\dot{n}}{n} \right)
 - 2\frac{\ddot{a}}{a} + \frac{\dot{b}}{b}\left( -2\frac{\dot{a}}{a}+\frac{\dot{n}}{n} \right)
 - \frac{\ddot{b}}{b} \Bigg\} - k\gamma_{ij},\\
G_{05}&=&3\left( \frac{n'}{n}\frac{\dot{a}}{a} + \frac{a'}{a}\frac{\dot{b}}{b} - \frac{\dot{a'}}{a} \right),\\
G_{55}&=&3\Bigg\{ \frac{a'}{a}\left( \frac{a'}{a}+\frac{n'}{n}
\right) - \frac{b^2}{n^2} \left( \frac{\dot{a}}{a} \left(
\frac{\dot{a}}{a}-\frac{\dot{n}}{n} \right) \right) -
k\frac{b^2}{a^2} \Bigg\},
\end{eqnarray}
where $\dot{\chi}=\frac{d\chi}{dt}$ and $\chi'=\frac{d\chi}{dy}$.
For having a well define geometry, the metric must be continuous
across the brane, but its derivative with respect to fifth
coordinate can be discontinues on the brane. This will entail the
presence of Dirac delta function in the second derivative metric
with respect to fifth coordinate. So according to {\cite{5}}, one
can obtain:
\begin{equation}
\frac{[a']}{b_0a_0}=-\frac{\kappa^2}{3}\rho_{b},
\end{equation}
\begin{equation}
\frac{[n']}{b_0n_0}=\frac{\kappa^2}{3}(3p_{b}+2\rho_{b}),
\end{equation}
the subscript 0 implies their value on the brane. By taking the
jump of (7) and using Eqs.(9) and (10), we derive the conservation
equation on the brane as follow:
\begin{equation}
\dot{\rho}_{b}+3\frac{\dot{a}_0}{a_0}(\rho_{b}+p_{b})=0.
\end{equation}
By imposing $Z_2$-symmetry on the average value of (8), one can
arrive at{\cite{5}}
\begin{equation}
\frac{\ddot{a}_0}{a_0}+\frac{\dot{a}_0^2}{a_0^2}=-\frac{\kappa^4}{36}
\rho_{b}(\rho_{b}+3p_{b}) -\frac{\kappa^2}{3b_0^2} P_{5}-
\frac{k}{a_0^2}.
\end{equation}
There is chosen $n_0=1$, this is possible by suitable time
transformation. Eq.(12) has important difference with standard
cosmology, there is quadratic energy density of the brane. This
term can be very important in early universe. One can rewrite 5-D
field equations (5) and (8) in the compact form as {\cite{5}}:
\begin{eqnarray}
\psi'&=&-\frac{2}{3}a'a^3\kappa^2\rho_{B},
\end{eqnarray}
\begin{eqnarray}
\dot{\psi}&=& \frac{2}{3}\dot{a}a^3\kappa^2P_{5},
\end{eqnarray}
where $\psi$ is a function of time and fifth coordinate,
\begin{equation}
\psi(t,y)\equiv\frac{(a'a)^2}{b^2}-\frac{(\dot{a}a)^2}{n^2}-ka^2.
\end{equation}
Now, we equate the time derivative of (13) with the fifth
coordinate derivative of (14), to get
\begin{equation}
a'\dot{\rho_{B}}+\dot{a}P'_5+(\rho_{B}+P_{5})\left(
\dot{a'}+3\frac{\dot{a}a'}{a} \right)=0,
\end{equation}
also, the constraint $\nabla_\alpha G^{\alpha0}=0$ gives
\begin{equation}
\dot{\rho_{B}}+3\frac{\dot{a}}{a}(\rho_{B}+P_{B})+\frac{\dot{b}}{b}(\rho_{B}+P_{5})=0.
\end{equation}
Eqs.(16)and(17) are the conservation relation on the bulk.\\
 By
imposing the $Z_2$-symmetry and junction condition (9) and the
result of average value of (15), one can obtain the generalized
Friedmann equation on the brane as follow:
\begin{equation}
\frac{\dot{a}_0^2}{a_0^2}=\frac{\kappa^4}{36}\rho_{b}^2-\frac{\psi_0(t)}{a_0^4}-\frac{k}{a_0^2}.
\end{equation}

\section{  Generalized Chaplygin Gas Model For Bulk }
In this work we choose a {\bf generalized chaplygin gas} for bulk, with the
equation of state as
\begin{equation}
P_B=P_5=A\rho_B-\frac{B}{\rho_B^\alpha},
\end{equation}
where $A$ and $B$ are positive constant, and $0<\alpha\leq1$. By
substituting this equation of state in the conservation equation
(17), one can obtain
\begin{equation}
\rho_B=\Bigg\{ \frac{1}{1+A} \left( \left( \frac{\rho_{B1}}{ba^3}
\right)^{(1+A)(1+\alpha)}+B \right) \Bigg\}^{\frac{1}{1+\alpha}},
\end{equation}
where $\rho _{B1}$ is an integration constant which can be
depends on $y$. According to the equation of state
\begin{eqnarray}
P_B=P_5&=&A\Bigg\{ \frac{1}{1+A} \left( \left(
\frac{\rho_{B1}}{ba^3} \right)^{(1+A)(1+\alpha)}+B \right)
\Bigg\}^{\frac{1}{1+\alpha}} \nonumber \\
 & & - \frac{B}{\Bigg\{ \frac{1}{1+A}
\left( \left( \frac{\rho_{B1}}{ba^3} \right)^{(1+A)(1+\alpha)}+B
\right) \Bigg\}^{\frac{\alpha}{1+\alpha}}}.
\end{eqnarray}
Since, we have assume that $\rho_B$ is independence on $y$, so
$\left( \frac{\rho_{B1}}{ba^3} \right)$ must be independence on
$y$. With the help of bulk conservation equations (16) and (17),
we arrive at this result that
\begin{equation}
\frac{\dot{b}}{b}=\frac{\dot{a}'}{a'}.
\end{equation}
One can integrate of (22) and find out a relation between
$b(t,y)$ and $a(t,y)$ as follow
\begin{equation}
b(t,y)=\beta(y)a'(t,y),
\end{equation}
where $\beta(y)$ is a integration constant which can be
dependence on $y$. Also, by using (22) in (7), one can realize
 that $n(t,y)$ is independence on $y$; namely
\begin{equation}
n'(t,y)=0.
\end{equation}
So, from the junction condition (10), we obtain an equation of
state for brane as
\begin{equation}
p_b=-\frac{2}{3}\rho_b.
\end{equation}
It is clearly seen that our assumptions impose a specific equation
of state on brane. From  the brane conservation equation (11), it
can be found out a relation between brane energy density,
$\rho_b$, and brane scale factor, $a_0(t)$, as follow
\begin{equation}
\rho_b =\frac{\rho_0}{a_0(t)},
\end{equation}
where $\rho_0$ is an integration constant. From another junction
condition (9) and also with the help of $Z_2$-symmetry we can
have a specific value for $\beta(y)$ in the $y=0$
\begin{equation}
\beta^{-1}(0)=-\frac{\kappa^2}{6}\rho_0.
\end{equation}
In the continuance of this work we suppose that, it is possible
to write $a(t,y)$ as a separate function of time and fifth
coordinate; namely
\begin{equation}
a(t,y)=f(y)a_0(t),
\end{equation}
where $f(y)$ is an arbitrary function of fifth coordinate, and
$a_0(t)$ is our brane scale factor, also it is only a function of
time. Because of $a(t,y=0)$ is brane scale factor, then as
$y\longrightarrow 0$, $f(y)$ must tends to $1$. From the (23),
$b(t,y)$ can be written as a separate function as well,
\begin{equation}
b(t,y)=\beta(y)f'(y)a_0(t)=g(y)b_0(t),
\end{equation}
for having nonzero value of $b(t,y=0)$, $f'(0)$ must not be
vanished. According to the (29), it is seen that, the time
{\bf behavior} of $b(t,y)$ is like to the $a(t,y)$. In here, we can
predict a function for $\rho_{B1}$ to give $\rho_B$ in which be
independence on $y$, as
\begin{equation}
\rho_{B1}=c\beta(y)f'(y)f^3(y),
\end{equation}
where $c$ is a coupling constant. Now, we rearrange the energy
density and pressure of bulk:
\begin{equation}
\rho_B=\Bigg\{ \frac{1}{1+A} \left(
\frac{c}{a_0^{4(1+A)(1+\alpha)}} +B \right)
\Bigg\}^{\frac{1}{1+\alpha}},
\end{equation}
and
\begin{eqnarray}
P_B=P_5&=&A\Bigg\{ \frac{1}{1+A} \left(
\frac{c}{a_0^{4(1+A)(1+\alpha)}} +B \right)
\Bigg\}^{\frac{1}{1+\alpha}} \nonumber \\
 & &- \frac{B}{\Bigg\{ \frac{1}{1+A} \left(
\frac{c}{a_0^{4(1+A)(1+\alpha)}} +B \right)
\Bigg\}^{\frac{\alpha}{1+\alpha}}}.
\end{eqnarray}
By using above assumption we arrive at this result that, for every
value of $y$ we have
\begin{equation}
\frac{\dot{a}(t,y)}{a(t,y)}=\frac{\dot{b}(t,y)}{b(t,y)}=\frac{\dot{a_0}(t)}{a_0(t)}.
\end{equation}
\par For investigation the evolution of universe, at first we need
to recognized $\psi(t,y)$. From (13), because $\rho_B$ is
independence on $y$, we have
\begin{equation}
\psi(t,y)=-\frac{\kappa^2}{6}\rho_B a^4 + \zeta(t),
\end{equation}
where $\zeta$ is a constant of integration which depends on time.
With equating the time derivative of (34) to (14), and with the
help of (33), we realize that $\dot{\zeta}(t)=0$, so $\zeta$ is a
constant value. Now, from (34), (12) and (18) one can rewrite the
evolution equations of brane as:
\begin{equation}
\dot{a_0}^2=\frac{\kappa^4}{36}\rho_0^2 +
\frac{\kappa^2}{6}\rho_B a_0^2 - \frac{\zeta}{a_0^2} - k,
\end{equation}
and
\begin{equation}
\ddot{a_0}=-\frac{\kappa^2}{3\beta^2(0)f'^2(0)}\frac{P_5}{a_0(t)}
- \frac{\kappa^2}{6}\rho_B a_0 + \frac{\zeta}{a_0^3}.
\end{equation}

According to the these two evolution equations, in early time
when $a_0\longrightarrow 0$, $\rho_B$ tends to infinite; also
$\dot{a_0}^2\longrightarrow +\infty$ and
$\ddot{a_0}\longrightarrow -\infty$. Then we have a decelerate
universe in early time. In late time as $a_0$ tends to infinite,
$\rho_B \longrightarrow \left( \frac{B}{1+A}
\right)^{\frac{1}{1+\alpha}}$. {\bf It means we arrive at a constant
value for bulk energy density and bulk pressure in late time. In
this era $\dot{a_0}^2\longrightarrow +\infty$ and $\ddot{a_0}
\longrightarrow -\infty$. So we have a decelerate universe in late
time.}

\section{Equivalence With The Cosmological Constant}
In this section we show that, with the help of some assumption
which can be seen in some works (e.g. \cite{5}), one can arrive
at equation of state of cosmological constant for bulk. We
suppose that the energy density and pressure of bulk which are
specified by $\rho_B$ and $P_B=P_5$ respectively, depend only on
time, and also the coefficient of fifth coordinate in metric;
namely $b^2(t,y)$, be independence on time. With the help of
conservation relations of bulk (16), (17) and above assumption one
can write
\begin{equation}
\dot{\rho_{B}} +(\rho_{B}+P_{5})\left(
\frac{\dot{a'}}{a'}+3\frac{\dot{a}}{a} \right)=0,
\end{equation}
and
\begin{equation}
\dot{\rho_{B}}+3\frac{\dot{a}}{a}(\rho_{B}+P_{B})=0.
\end{equation}
By using (38) in (37), we have
\begin{equation}
(\rho_{B}+P_5)\frac{\dot{a'}}{a'}=0.
\end{equation}
This relation is satisfied in every time. Here, we have two situation:

\begin{itemize}
\item{\bf  If $(\rho_{B}+P_5)=0$.}\\
 then we arrive at this result that
\begin{equation}
P_5=P_B=-\rho_{B}.
\end{equation}
This is the equation of state of cosmological constant. By
substituting this relation in (38) we realize that, $\rho_b$ is
independence on time as well, and we get a constant value for
bulk energy density.

{ \item{\bf If $\dot{a}'=0$.}\\
In this situation with the help of (7), we realized that $n'=0$,
so from junction condition (10), we get a specific equation of
state for brane; namely
\begin{equation}
p_b=-\frac{2}{3}\rho_b.
\end{equation}
Whereas, $\dot{a}'=0$ therefore
\begin{equation}
a(t,y)=X(t)+Y(y),
\end{equation}
this form is an unusual form for coefficient of metric. From
conservation equation (38), for {\bf generalized chaplygin gas} equation of state,
we have
\begin{equation}
\rho_B=\Bigg\{ \frac{1}{1+A} \left( \left( \frac{\rho_1}{a}
\right)^{3(1+A)(1+\alpha)}+B \right) \Bigg\}^{\frac{1}{1+\alpha}},
\end{equation}
where $\rho_1$ is a constant of integration which can be depended
on $y$. Since, $\rho_B$ is independence on $y$, then
$\left(\frac{\rho_1}{a} \right)$ must be independence on $y$;
namely
\begin{equation}
\frac{d}{dy}\left(\frac{\rho_1(y)}{a(t,y)} \right)=0.
\end{equation}
From this relation, we should find out a function for $\rho_1$.
With the help of (44), we arrive at
$$\frac{\rho_1'}{\rho_1}=\frac{Y'}{Y+X}. $$ By some working, we
find out that, $\rho_1=Y(y)$. Now, derivative of
$\left(\frac{\rho_1}{a} \right)$ with respect to $y$ shows that
$Y'=0$. From (9), one can realize that, it means $\rho_b=0$,
while this result is incompatible with the experiment. So, this
case is not suitable.}

\end{itemize}

\section{Conclusion}
Finally, we see that in early time the bulk energy density and
bulk pressure are infinite; however , in late time they arrive at
a constant value. Also, it was shown that, our assumption imposed
a specific equation of state on the brane, and we can not give
any other equation of state to brane. {  We have shown that in this
model we have a decelerate universe in early time and late time.}
Also, it was seen that, when we supposed that, as well as
previous assumption, the coefficient of fifth coordinate in
metric, namely $b(t,y)$, be independence on time, there is
equation of state of cosmological constant on the bulk. So, we
have a constant value for bulk energy density.


\begin{thebibliography}{99}

\bibitem{1} D. Bailin, A. Love, Rep. Prog. Phys. 50 (1987), 1087.

\bibitem{2} L. Randall and R. Sundrum, Phys. Rev. Lett. {\bf 83}, (1999), 3370.

\bibitem{3} L. Randall and R. Sundrum, Phys. Rev. Lett. {\bf 83}, (1999), 4690.

\bibitem{4} T. Shiromizu, K. Maeda and M. Sasaki, Phys. Rev. D {\bf 62}, (2000), 024012.

\bibitem{5} P. Binetruy, C. Deffayet and D. Langlois, Nucl. Phys. B {\bf 565}, (2000), 269;\\
P. Binetruy, C. Deffayet, U. Ellwanger and D. Langlois, Phys.
Lett. B {\bf 477}, (2000), 285.


\bibitem{6} R. Maartens, Living Rev. Rel. {\bf 7}, (2004), 7 .

\bibitem{7} E. Papantonopoulos, Lect. Notes Phys. {\bf 592}, (2002), 458.

\bibitem{8} P. Brax, C. van de Bruck, Class. Quantum Gravity {\bf 20},
(2003), R201-R232.



\bibitem{9} C. Csaki, M. Graesser, C. F. Kolda and J. Terning, Phys. Lett. B {\bf 462},
 (1999) 34; \\
 M. R. Setare, Phys. Lett. B {\bf 642}, (2006), 421;\\
 M. R. Setare and E. N. Saridakis, JCAP {\bf 0903},
(2009), 002;\\
J. M. Cline, C. Grojean and G. Servant, Phys. Rev. Lett. {\bf 83},
(1999), 4245.

\bibitem{10} P. Kraus, JHEP {\bf 9912}, (1999), 011.

\bibitem{11} D. Ida, JHEP {\bf 0009}, (2000), 014.

\bibitem{12} R. Maartens (private communication); A.Mazumdar,
R. N. Mohapatra, and A. Perez-Lorenzana, JCAP {\bf 0406}, (2004), 004 ;\\
C. Bogdanos and K. Tamvakis, Phys. Lett. B {\bf 646}, (2007), 39.

\bibitem{13} M. R. Setare, Phys. Lett. B {\bf 654}, (2007), 1-6; \\
M. R. Setare, Phys. Lett. B {\bf 648}, (2007), 329.



\end{thebibliography}
\end{document}